\documentclass[journal,twocolumn]{IEEEtran}
\usepackage{cite,amsmath,amssymb,amsfonts,algorithmic}
\usepackage{graphicx,textcomp,multirow,booktabs,subfigure,hyperref}

\begin{document}
\title{Effective and Efficient Intracortical Brain Signal Decoding with Spiking Neural Networks}
\author{Haotian Fu, Peng Zhang, Song Yang, Herui Zhang, Ziwei Wang and Dongrui Wu, \IEEEmembership{Fellow, IEEE}
\thanks{H. Fu, S. Yang, H. Zhang, Z. Wang and D. Wu are with the Key Laboratory of the Ministry of Education for Image Processing and Intelligent Control, School of Artificial Intelligence and Automation, Huazhong University of Science and Technology, Wuhan 430074, China.}
\thanks{P. Zhang is with Department of Biomedical Engineering, College of Life Science and Technology, Huazhong University of Science and Technology, Wuhan 430074, China}
\thanks{Dongrui Wu is the corresponding author (drwu09@gmail.com).}}

\maketitle

\begin{abstract}
A brain-computer interface (BCI) facilitates direct interaction between the brain and external devices. To concurrently achieve high decoding accuracy and low energy consumption in invasive BCIs, we propose a novel spiking neural network (SNN) framework incorporating local synaptic stabilization (LSS) and channel-wise attention (CA), termed LSS-CA-SNN. LSS optimizes neuronal membrane potential dynamics, boosting classification performance, while CA refines neuronal activation, effectively reducing energy consumption. Furthermore, we introduce SpikeDrop, a data augmentation strategy designed to expand the training dataset thus enhancing model generalizability. Experiments on invasive spiking datasets recorded from two rhesus macaques demonstrated that LSS-CA-SNN surpassed state-of-the-art artificial neural networks (ANNs) in both decoding accuracy and energy efficiency, achieving 0.80-3.87\% performance gains and 14.78-43.86 times energy saving. This study highlights the potential of LSS-CA-SNN and SpikeDrop in advancing invasive BCI applications.
\end{abstract}


\section{Introduction} \label{sec:introduction}

\begin{figure}[htbp]     \centering
    \includegraphics[width=\linewidth,clip]{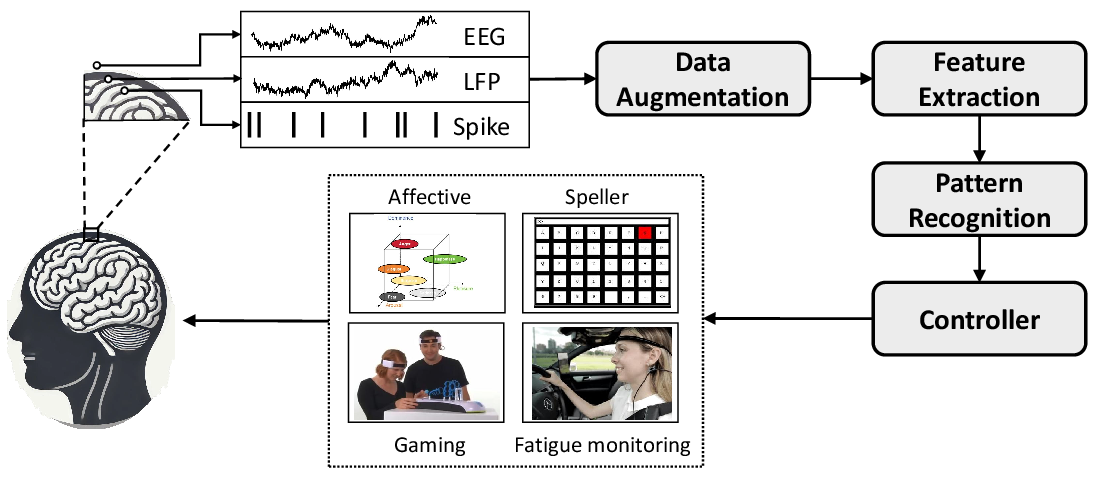}
    \caption{A closed-loop BCI system.}     \label{fig1pipeline}
\end{figure}

Brain-computer interfaces (BCIs) facilitate direct interaction between the brain and external devices, bypassing conventional neuromuscular pathways~\cite{Lance2012}. Originally designed to assist individuals with severe motor impairments~\cite{Pfurtscheller2008}, BCIs have progressively expanded their scope to encompass applications for able-bodied users, such as emotion recognition~\cite{drwuPIEEE2023}, text input~\cite{Chen2015a}, gaming~\cite{Marshall2013}, and driver fatigue detection~\cite{drwuFWET2019}, as depicted in Fig.~\ref{fig1pipeline}.

BCIs are generally categorized into three types based on electrode placement: non-invasive, semi-invasive, and invasive~\cite{Wu2020}. Non-invasive BCIs predominantly rely on electroencephalography (EEG) signals, while semi-invasive BCIs utilize electrocorticography (ECoG) signals, which are recorded from electrodes implanted beneath the skull but external to brain tissue. Invasive BCIs employ microelectrodes implanted directly into brain tissue, achieving superior spatial and temporal resolution~\cite{Zhang2020}. These electrodes capture action potentials or ``spikes" from individual neurons, enabling precise control of prosthetic devices and communication systems~\cite{Bouton2016}. However, invasive BCIs face significant challenges, including risks of infection, scar tissue development, and long-term instability arising from the brain’s immune response~\cite{Gallego2020}.

Artificial neural networks (ANNs) have demonstrated significant success in BCI applications, excelling in areas such as accurate decoding~\cite{Wu2022}, privacy preservation~\cite{Zhang2022}, and adversarial robustness~\cite{Wu2023}. However, these advancements come with the drawback of substantial energy consumption. In mobile BCI systems, high energy consumption diminishes device operational longevity, while in invasive BCIs, excessive energy dissipation poses risks of neuronal damage due to heat generation.

To achieve both high decoding accuracy and low energy consumption in invasive BCIs, we propose a local synaptic stabilization and channel-wise attention-based spiking neural network (LSS-CA-SNN). Spiking neural networks (SNNs) are leveraged for their ability to process sparse spiking data in an event-driven, energy-efficient manner. The local synaptic stabilization (LSS) mechanism enhances decoding accuracy by stabilizing synaptic responses and optimizing neuronal membrane potentials, while the channel-wise attention (CA) module focuses selectively on critical neural activities, suppressing irrelevant noise to improve computational efficiency.

Moreover, we introduce SpikeDrop, a spiking data augmentation technique to expand the training dataset thus enhancing model generalizability. By randomly masking spiking data from diverse perspectives, SpikeDrop simulates a variety of spiking patterns, effectively increasing the diversity of training samples. When integrated with SpikeDrop, LSS-CA-SNN further boosts performance. To the best of our knowledge, this represents the first data augmentation approach specifically tailored for spiking data.

The remainder of this paper is structured as follows: Section~\ref{sec:related works} reviews related literature, Section~\ref{sec:methods} details the proposed LSS-CA-SNN framework and SpikeDrop method, Section~\ref{sec:experiment} presents experimental results and corresponding analyses, and Section~\ref{sec:conclusion} concludes the study.

\section{Related Works} \label{sec:related works}

This section introduces related works on ANNs for noon-invasive BCI decoding, SNNs for invasive BCI decoding, and data augmentation.

\subsection{ANNs for Non-Invasive BCI Decoding}

In recent years, deep learning has significantly advanced the applications of ANNs in non-invasive BCIs. These approaches typically employ convolutional layers to extract discriminative features from the spatial, temporal, and frequency domains of EEG signals~\cite{Wu2022}.

DeepConvNet and ShallowConvNet are among the seminal contributions in this field~\cite{Schirrmeister2017}. DeepConvNet offers a general-purpose framework for non-invasive BCI decoding, while ShallowConvNet integrates temporal and spatial convolutional layers to directly capture task-relevant EEG features. Building upon these efforts, Lawhern et al. introduced EEGNet\cite{Lawhern2018}, a compact architecture that combines temporal convolutions with spatial depthwise convolutions, demonstrating effectiveness in various EEG-based BCI paradigms. More recent advancements include EEGConformer~\cite{Song2022}, which incorporates attention mechanisms to enhance the decoding capabilities of convolutional neural network (CNN) for EEG signals.

However, the significant differences between continuous EEG signals and discrete binary spiking signals present substantial challenges for adapting EEG-specific models to process spiking data efficiently. Hence, it is necessary to develop SNN-based models for efficient spiking decoding.

\subsection{SNNs for Invasive BCI Decoding}

SNNs, as the third generation of neural networks, are an energy-efficient alternative to ANNs. Unlike ANNs, which depend on continuous signal transmission, SNNs communicate using binary spiking signals, aligning more closely with biological neural systems. The event-driven and asynchronous nature of SNNs significantly reduces the computational and memory overhead, especially when deployed on neuromorphic hardware like TrueNorth~\cite{Merolla2014}, Loihi~\cite{Davies2018}, and Tianjic~\cite{Pei2019}.

The non-differentiability of spiking neurons poses a challenge to the development of SNNs. Early methods convert pre-trained ANNs into SNNs by substituting the activation functions with spiking neurons. However, these approaches often require more time steps and do not perform as well as ANNs~\cite{Diehl2015}. Spatio-temporal backpropagation (STBP) uses surrogate gradients to approximate activities of spiking neurons, allowing SNNs to be trained end-to-end via backpropagation~\cite{Wu2018}. This greatly improves the performance of SNNs. Other research focuses on making spiking neurons more adaptable by introducing learnable parameters, e.g., parameterized leaky integrate-and-fire (PLIF) neuron adopts learnable membrane time constant to boost neuron performance~\cite{Fang2021}, and DIET-SNN~\cite{Rathi2021} combines learnable membrane leakage and firing threshold to optimize SNNs efficiency and adaptability for complex tasks.

SNNs have shown promise in invasive BCIs. Dethier et al.~\cite{Dethier2011} implemented an SNN-based decoder to predict the arm movement speed of rhesus monkeys, achieving performance comparable to traditional floating-point techniques. Liao et al.~\cite{Liao2022} proposed an energy-efficient SNN for decoding finger velocity in invasive BCIs, demonstrating that SNN requires significantly fewer computational operations and memory accesses than ANN at similar accuracies. Feng et al.~\cite{Feng2023} developed an ANN-to-SNN conversion method for EEGNet, optimizing it for motor imagery datasets and achieving nearly 10x inference efficiency on neuromorphic hardware.

\subsection{Data Augmentation}

Data augmentation is a critical technique for enhancing dataset diversity and volume by applying systematic transformations to existing samples. It plays a pivotal role in mitigating overfitting and enhancing the generalization capabilities of machine learning models.

In BCIs, data augmentation has become instrumental in improving model generalization, mitigating data variability, and tackling alignment challenges. Most EEG-based data augmentation techniques have explored transformations within the time, frequency, and spatial domains. In the time domain, Wang et al.~\cite{Wang2018} added Gaussian white noise to the original EEG signals, and Mohsenvand et al.~\cite{Mohsenvand2020} set random EEG segments to zero. In the frequency domain, Schwabedal et al.~\cite{Schwabedal2018} randomized the phases of Fourier transform on all channels. In the spatial domain, Wang et al.~\cite{Wang2024} proposed channel reflection, which swaps channels between the left and right hemispheres, and Saeed et al.~\cite{Saeed2021} set values of randomly selected channels to zero or permuted them.

While these methods have proven successful for EEG signals, their applicability to spiking data is limited due to their discrete and event-driven nature.

For data augmentation of event-based datasets. Neuromorphic data augmentation (NDA)~\cite{Li2022} uses a series of transformations to increase the pattern of event data, including rolling, cutting, shearing, and rotation. Eventmix~\cite{Shen2023} develops a three-dimensional mask based on the spatiotemporal characteristics of event data.

To the best of our knowledge, no prior work has specifically investigated data augmentation for invasive spiking signals, leaving this area largely unexplored.

\section{Methods} \label{sec:methods}

\begin{figure*}[htbp] 	\centering
	\includegraphics[width=\linewidth,clip]{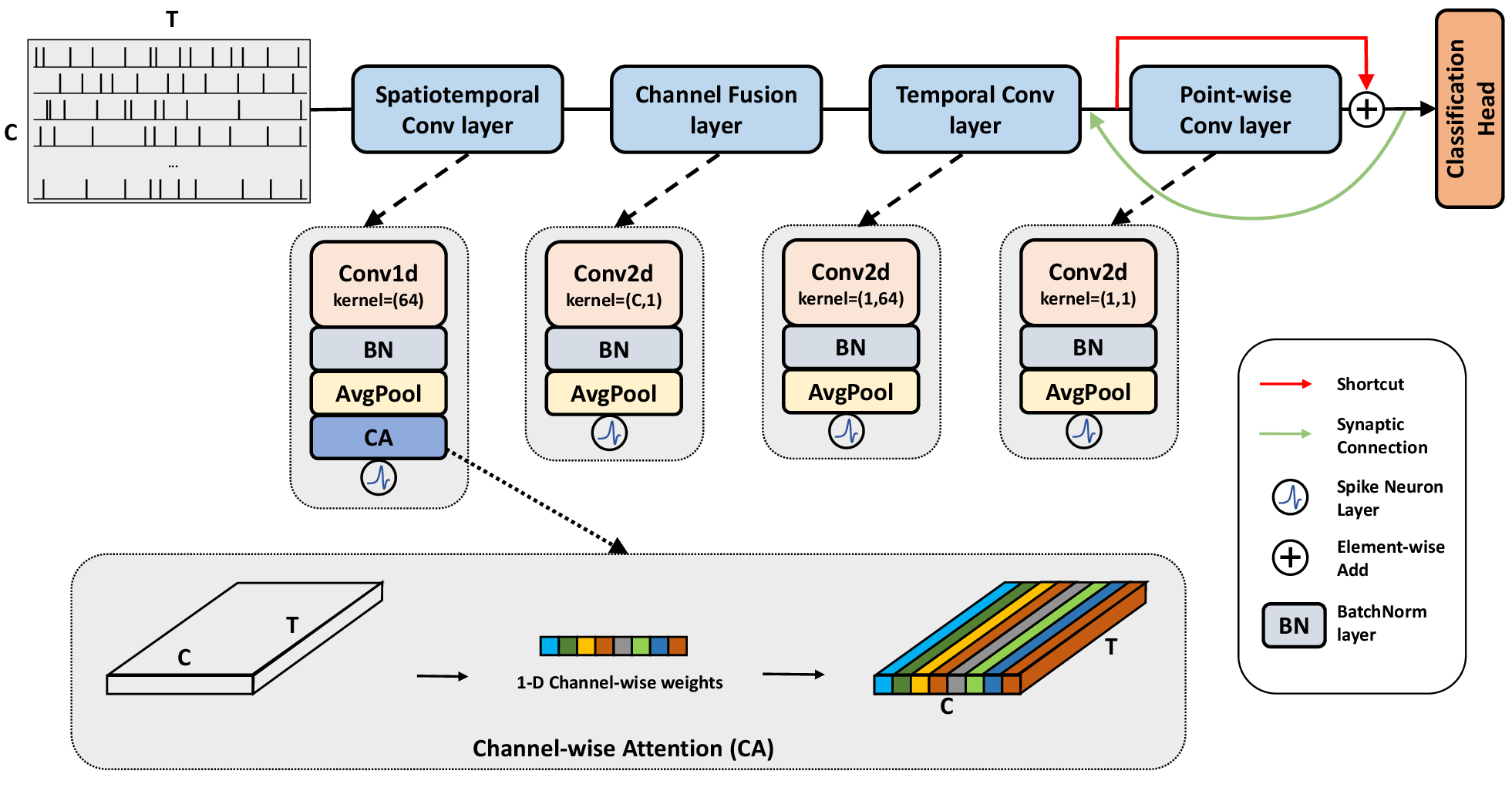}
	\caption{Our proposed LSS-CA-SNN, which includes four different convolutional layers to extract both spatial and temporal features from spiking signals.} 	\label{fig3net}
\end{figure*}

This section introduces our proposed LSS-CA-SNN, as shown in Fig.~\ref{fig3net}.

\subsection{Preliminaries of SNN}

Spiking neurons are the fundamental computational units of SNNs, communicating through spikes encoded as binary activations. The PLIF model is widely adopted due to its balance between biological plausibility and computational complexity.

The PLIF neuron is characterized by a differential equation for membrane potential, which can be efficiently solved for large-scale SNN simulations:
\begin{align}
    \tau \frac{\mathrm{d} u(t)}{\mathrm{d} t}=-u(t)+I(t),
\end{align}
where $\tau$ (initialized to 2) is the learnable time constant, and $u(t)$ and $I(t)$ the membrane potential and input current, respectively.

For numerical simulation of the PLIF neuron layer in an SNN, a discrete-time parametric dynamical version is considered, which incorporates a threshold-triggered firing mechanism and post-firing reset of the membrane potential. The dynamics of various spiking neurons can be described as:
\begin{align}
\left\{\begin{array}{l}
\boldsymbol{U}^{t, n}=
    \boldsymbol{H}^{t-1,n}+\frac{1}{\tau}(-(\boldsymbol{H}^{t-1, n}-V_{\text{reset}})+\boldsymbol{X}^{t, n}), \\
\boldsymbol{S}^{t, n}=\text{Heav}\left(\boldsymbol{U}^{t, n}-u_{\text{th}}\right), \\
\boldsymbol{H}^{t, n}=V_{\text{reset}} \boldsymbol{S}^{t, n}+\left(\beta \boldsymbol{U}^{t, n}\right) \odot\left(\mathbf{1}-\boldsymbol{S}^{t, n}\right),
\end{array}\right.
\end{align}
where $t$ and $n$ denote the indices of time step and layer, respectively, and $\boldsymbol{U}^{t,n}$ and $\boldsymbol{H}^{t-1,n}$ respectively the membrane potential after neuronal dynamics and after the trigger of a spike. $\boldsymbol{X}^{t,n}$ is the external input, $u_{\text{th}}$ is the threshold, $\text{Heav}(x)$ is the Heaviside function, $V_{\text{reset}}$ is the reset potential, $\beta = e^{-\frac{\mathrm{t}}{\tau}}$ is the decay factor, and $\odot$ is the element-wise multiplication.

\begin{figure}[htbp] 	\centering
	\includegraphics[width=\linewidth,clip]{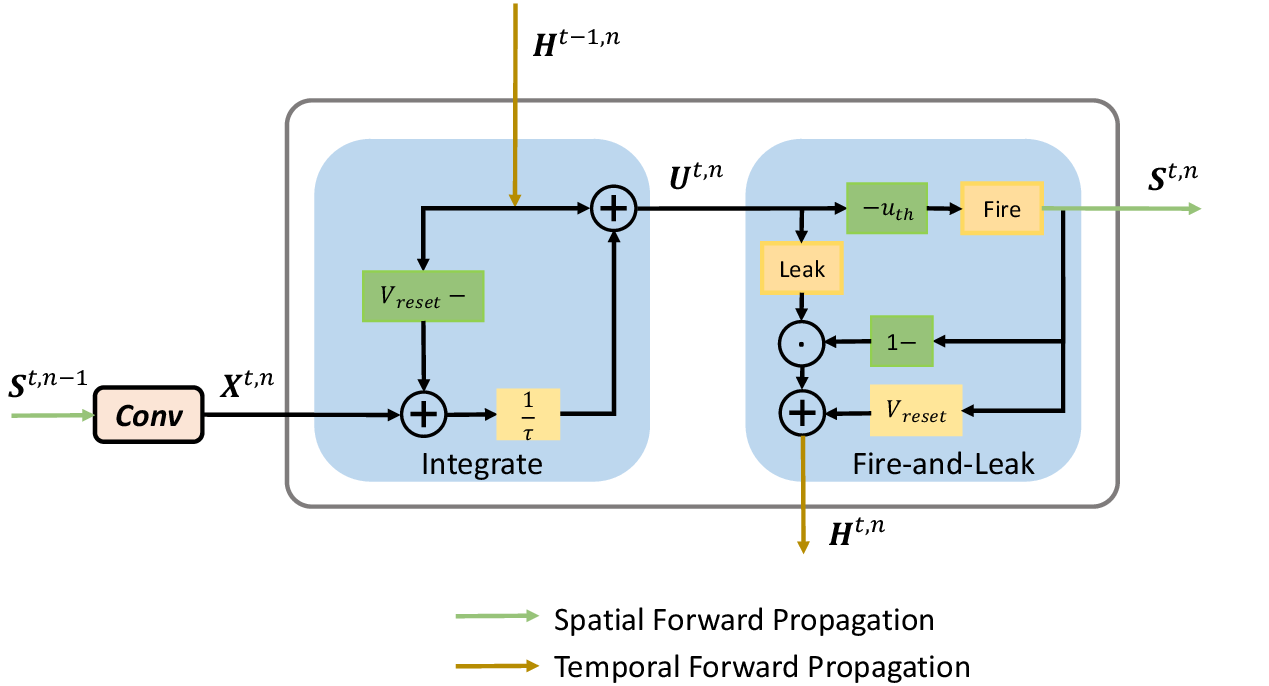}
	\caption{A convolution-based PLIF-SNN layer, including convolution, membrane potential integration, spiking activity, and synaptic leak.} \label{fig2LIF}
\end{figure}

Fig.~\ref{fig2LIF} illustrates the fire-and-leak mechanism in SNN, where the PLIF-SNN layer consists of a convolutional module followed by a PLIF neuron module. The Conv module processes the raw input spike tensor $\boldsymbol{S}^{t, n-1}$, extracting spatial features from the input. Then, the PLIF neuron module takes the spatial features  $\boldsymbol{X}^{t, n}$ and integrates them with the temporal input $\boldsymbol{H}^{t-1, n}$ to calculate the membrane potential $\boldsymbol{U}^{t, n}$. If the membrane potential $\boldsymbol{U}^{t, n}$ exceeds the predefined threshold $u_{\text{th}}$, the neuron fires a spike, and the membrane potential is reset to $V_{\text{reset}}$.

Direct SNN training is challenging due to the non-differentiability of the spiking activation function. Surrogate gradients are used for spatio-temporal backpropagation, allowing efficient learning despite the discrete nature of spikes~\cite{Wu2018}:
\begin{align}
     \frac{\partial \boldsymbol{S}^{t}}{\partial \boldsymbol{U}^{t}}=\frac{1}{a} \operatorname{sign}\left(\left|\boldsymbol{U}^{t}-u_{\text{th}}\right|<\frac{a}{2}\right) .
\end{align}
$a=4$ was used in our experiments.

\subsection{SNN with Synaptic Stabilization and Attention}

To effectively capture spatiotemporal dynamics of spiking data, we propose LSS-CA-SNN, which consists of three main modules: spatiotemporal feature extraction, channel fusion with multilayer spatiotemporal convolutions, and classifier, as shown in Fig.~\ref{fig3net}. LSS-CA-SNN integrates synaptic stabilization and attention mechanism.

\subsubsection{Spatiotemporal feature extraction}

The raw spiking signal from a single trial can be represented as a 2D matrix $X \in \mathbb{B}^{C \times T}$, where $\mathbb{B}=\{0,1\}$, $C$ is the number of channels, and $T$ the number of time points. Our module starts with a 1D convolutional layer with kernel size 64, designed to effectively capture temporal features within each individual channel. The initial 1D convolution facilitates comprehensive extraction of information across channels, ensuring that inter-channel dependencies are well-represented. Then, batch normalization is applied to stabilize the input distribution, followed by a channel-wise attention module that assigns appropriate weights to each channel, enhancing relevant features. The output is then passed through an average pooling layer to reduce the feature dimensionality, followed by a PLIF-based spiking activation function, which introduces nonlinearity and enables the model to process discrete spiking information.

\subsubsection{Channel fusion with multilayer spatiotemporal convolutions}

This module enhances the network's ability to capture cross-channel interactions within the input data. It includes three layers: channel fusion, temporal convolution, and refine convolution.

We start by fusing channel data using a 2D convolution operation with kernel size $(C, 1)$, which encourages feature interactions across channels, inspired by complex-valued convolutional fusion~\cite{Wang2024b}. Batch normalization and average pooling are then applied to normalize data and reduce the temporal dimensionality of the feature maps.

The data are then processed by two convolutional layers. The first uses a $(1, 64)$ kernel to capture sequential features across the time dimension, allowing the network to extract relevant patterns over time. The second convolutional layer uses a $(1, 1)$ kernel to further refine the mutual mapping between the feature maps, enhancing the model's ability to focus on essential characteristics without altering the spatial structure. This combination of sequential feature extraction and refined mapping ensures that the network captures complex patterns while maintaining computational efficiency.

To overcome the vanishing gradient issue in deeper layers, especially in SNNs, we introduce residual connections in the last layer of LSS-CA-SNN, ensuring the effective utilization and propagation of deep features. Inspired by biological neuronal connections, we incorporate a local synaptic stabilization mechanism in the fourth layer, which improves the learning efficiency and performance by optimizing the membrane potential of deep neurons.

\subsubsection{Classifier}

After hierarchical feature extraction and convolutional operations, the network aggregates the extracted features using global average pooling. This operation reduces each feature map to a single value, resulting in a fixed-dimensionality feature vector aligned with the number of target classes, reducing the model's complexity while retaining essential information.

To mitigate overfitting, a dropout layer is introduced, which encourages the model to rely on multiple pathways through the network and enhances its generalization ability. Finally, the processed features are passed through a fully connected layer for classification.

\subsection{SpikeDrop Data Augmentation}

\begin{figure}[htbp]    \centering
    \includegraphics[width=\linewidth,clip]{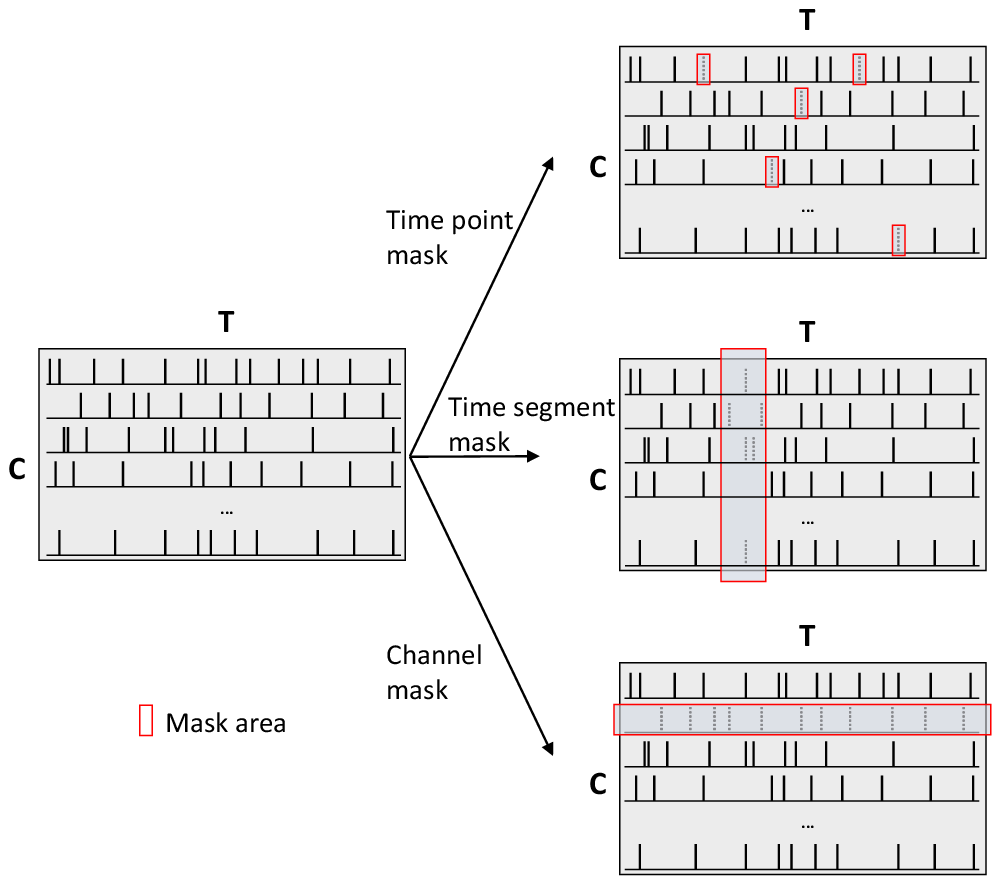}
    \caption{SpikeDrop augmentation process, showcasing the application of masks from various perspectives to manipulate the spiking data. }     \label{fig4aug}
\end{figure}

To enhance the generalization of LSS-CA-SNN, we develop a novel spiking data augmentation technique called SpikeDrop, as shown in Fig.~\ref{fig4aug}. SpikeDrop performs random masking across time points, time segments, and channels, to stabilize the training and improve generalization.

Assume the training data have dimensionality $N \times C \times T$, where $N$ is the number of samples, $C$ the number of channels, and $T$ the time points. SpikeDrop selects $M$ samples from the dataset for multi-perspective random masking: time point masking, time segment masking, and channel masking.

Time point masking uses a binary mask matrix $M_{t} \in \mathbb{B}^{C \times T}$, initialized to all ones, where each element has a probability $p_{t}=0.1$ of being masked to zero if a spike is present, thus simulating the absence of a spike:
\begin{align*}
M_{t}[c, t]=\left\{\begin{array}{ll}
0, & \text { if } X[c, t]=1 \text { and with probability } p_{t} \\
1, & \text { otherwise}
\end{array}\right..
\end{align*}
The masked data $X'$ are represented as $X' = X \odot M_{t}$.

For time segment masking, we define a mask matrix $M_{s} \in \mathbb{B}^{1 \times T}$ that selects a random time segment $[t_s,t_s+l]$ and masks all channels within this segment, with probability $p_{s}=0.05$:
\begin{align*}
M_{s}[t]=\left\{\begin{array}{ll}
0, & \text { if } t \in [t_s,t_s+l]\text{ and } \text {with probability } p_{s} \\
1, & \text { otherwise }
\end{array}\right..
\end{align*}
The masked data are represented as $X' = X \odot M_{s}$, leveraging the broadcasting mechanism.

Channel masking applies a mask matrix $M_{c} \in \mathbb{B}^{C \times 1}$, where a channel $c_i$ is masked with probability $p_c=0.1$:
\begin{align*}
M_{c}[c_i]=\left\{\begin{array}{ll}
0, & \text { if } \text {with probability } p_{c} \\
1, & \text { otherwise}
\end{array}\right..
\end{align*}
The masked data are computed as $X' = X \odot M_{c}$, again utilizing the broadcasting mechanism.

This systematic data augmentation approach ensures that the model is exposed to a diverse set of input patterns, crucial for its generalization.

\section{Experiments and Results} \label{sec:experiment}

This section presents experiment results to demonstrate the effectiveness and efficiency of our proposed approaches. The code is available on GitHub\footnotemark.
\footnotetext{\url{https://github.com/1439278026/LSS-CA-SNN}}

\subsection{Dataset Description}

We utilized data from two adult male rhesus macaques, Monkey B and Monkey H, who underwent distinct training protocols for similar spatial reaching and grasping tasks~\cite{Zhang2018}.

\begin{figure*}[hbtp]    \centering
    \subfigure[]{
        \includegraphics[width=0.3\linewidth,clip]{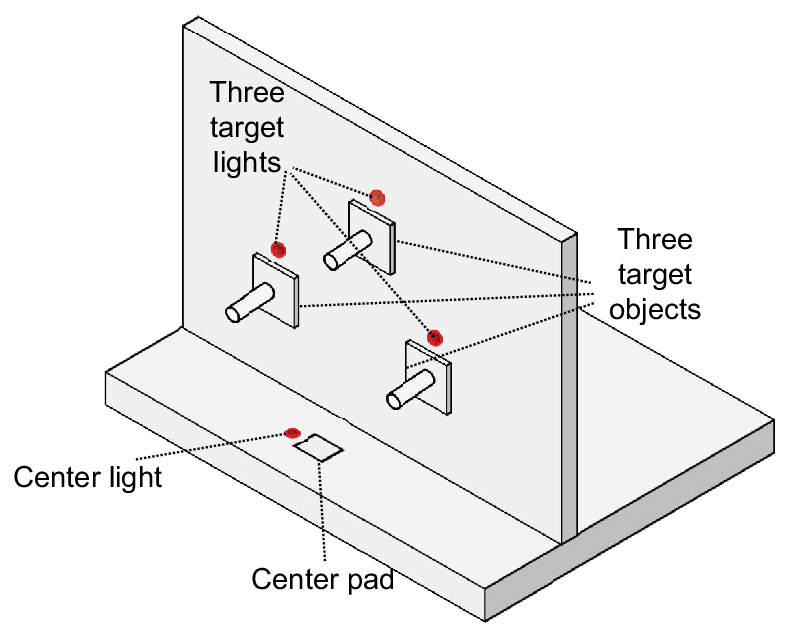}	        \label{fig5a}    }\qquad
    \subfigure[]{
        \includegraphics[width=0.3\linewidth,clip]{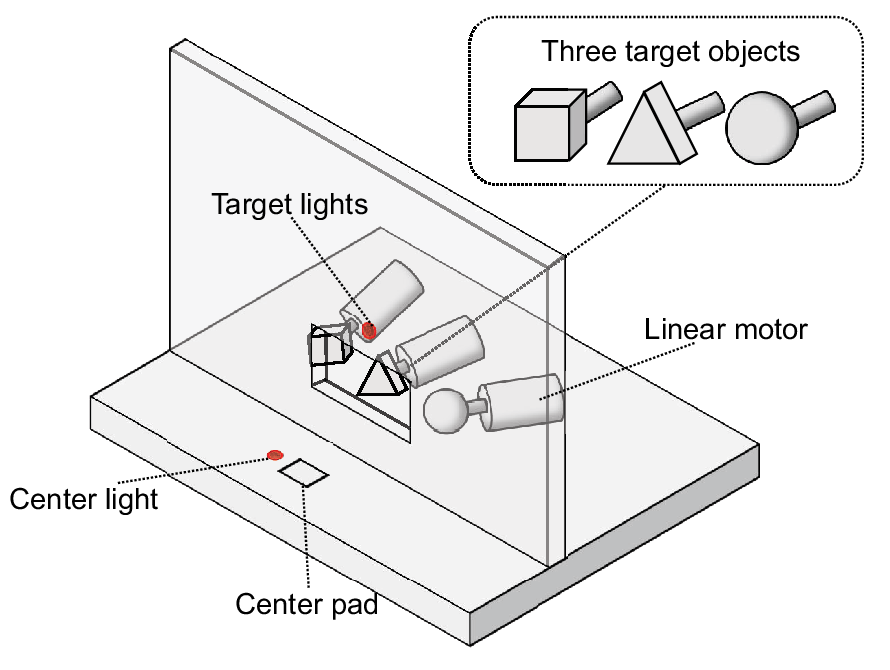}        \label{fig5b}    }
    \subfigure[]{
        \includegraphics[width=.8\linewidth,clip]{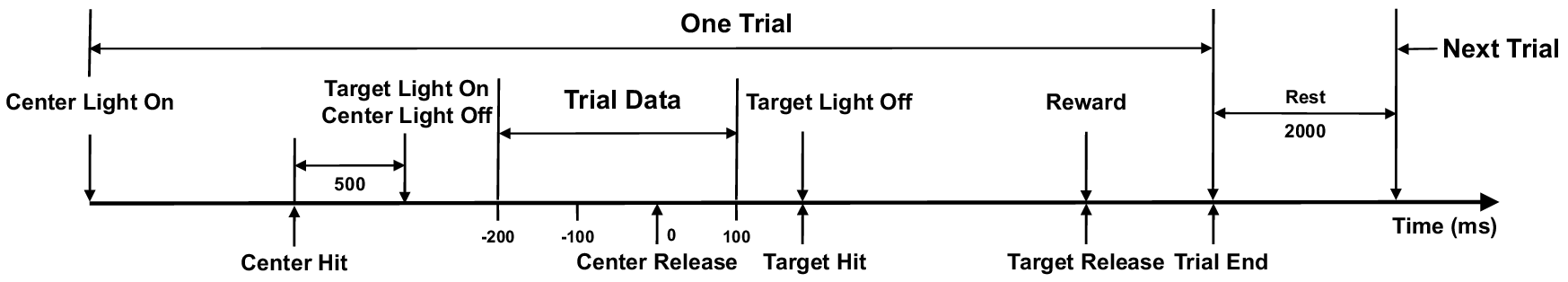}        \label{fig5c}    }
    \subfigure[]{
        \includegraphics[width=.8\linewidth,clip]{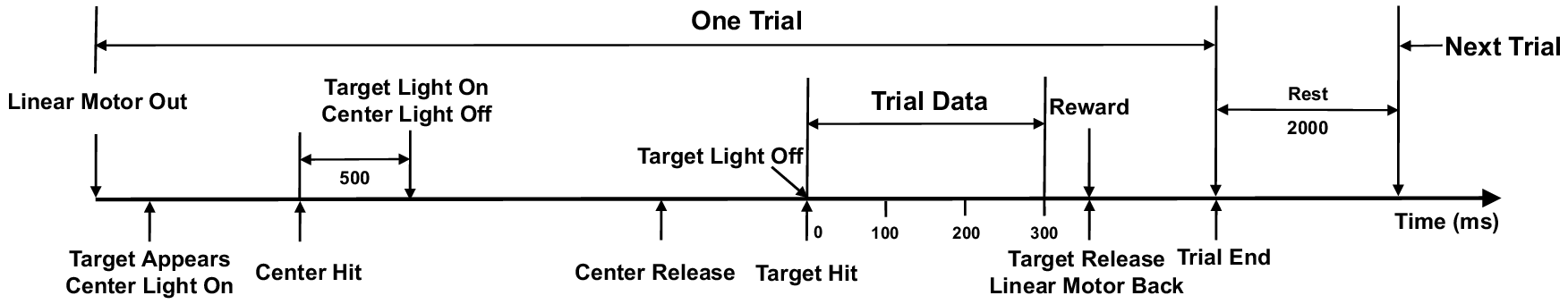}         \label{fig5d}     }
    \caption{The data collection device in (a) motor paradigm and (b) sensory paradigm, and the experimental process in (c) motor paradigm and (d) sensory paradigm.}     \label{fig5}
\end{figure*}

The experimental devices are illustrated in Figs.~\ref{fig5a}-\ref{fig5b}. Monkey B was trained in a ``motor paradigm", focusing on motor decoding by reaching and grasping objects at different positions. In contrast, Monkey H was trained in a ``sensory paradigm", grasping objects of various shapes at the same position to focus on sensory decoding.

The task sequences are illustrated in Figs.~\ref{fig5c}-\ref{fig5d}. Trials of Monkey B began with the activation of a center light, while Monkey H initiated trials with linear motor contacted with the target object going forward and the center light was on when the target object reached the fixed position. The monkeys were indicated by the center light to touch the center pad. After touching the center pad for approximately 500 ms, the center light turned off, and a single target light went on, prompting the monkey to grasp the target fully. ``Center release" was defined as the monkey's hand leaving the center pad, and ``target hit" as touching the target object. For Monkey H, the linear motor retracted upon target release. Successful trials were rewarded with water drops. If the monkey interrupted the hold or grasped incorrectly, the trial was aborted.

Neural electrodes were implanted in brain regions related to movement and sensory processing. Monkey B had two 32-channel Utah arrays in M1 and S1, and a 16-channel floating microelectrode array in the posterior parietal cortex. Monkey H had four 32-channel floating microelectrode arrays in M1, S1, and posterior parietal cortex (two). A total of 80 functional channels of Monkey B and 66 channels of Monkey H were used for neural signal decoding.

Data were recorded using a 128-channel OmniPlex system, with neural signals amplified, 250-6000Hz band-pass filtered, and sampled at 40 kHz. Spikes were extracted using a threshold detection method set at -4.5 times of the root mean square of the spike band for each channel. We used threshold crossings, providing stable decoding performance over time. The use of the OmniPlex system and threshold detection method were consistent with standard practices in neural signal recording, ensuring high-quality data for our analysis.

For the motor paradigm, we selected a $[-200, 100]$ ms time window around the center release event to decode the motion intention. For the sensory paradigm, a $[0, 300]$ ms time window after the target hit event was used to decode hand posture. We then reduced the trial length from 12000 to 100 using max pooling while maintaining the spike format. The dataset included 14 sessions for Monkey B and 12 sessions for Monkey H, with balanced classes. Table~\ref{tablesummary} summarizes the dataset characteristics.

\begin{table}[htbp]
\caption{Summary of the two datasets.} \centering\setlength{\tabcolsep}{1mm}
\begin{tabular}{cccccc}
\toprule[1pt]
 & \begin{tabular}[c]{@{}c@{}}Number of\\ Sessions\end{tabular} & \begin{tabular}[c]{@{}c@{}}Number of\\ Channels\end{tabular} & \begin{tabular}[c]{@{}c@{}}Number of\\ Classes\end{tabular} & Task Type & Paradigm \\ \midrule
Monkey B & 14 & 80 & 3 & Direction & Motor\\
Monkey H & 12 & 66 & 3 & Shape & Sensory\\ \bottomrule[1pt]
\label{tablesummary}
\end{tabular}
\end{table}

\subsection{Experiment Settings}

To evaluate the performance of LSS-CA-SNN and SpikeDrop, we designed two experimental settings different in the amount of training data used:
\begin{enumerate}
\item {\it Cross-session unsupervised transfer}, where the training dataset consisted of only labeled data from the source sessions, without any labeled data from the target session. This scenario simulated a real-world situation where the model must generalize from some sessions to an entirely new, unseen session.
\item {\it Cross-session supervised transfer}, where the training set included all labeled data from the source sessions, and a smaller set of labeled data from the target session. This setup examined the model's ability to leverage a limited amount of calibration data from the target session to improve its performance.
\end{enumerate}

For each paradigm, each session was utilized as the target session once, and the remaining sessions as the source sessions. This approach ensured a comprehensive evaluation across different data distributions. To ensure the reliability and reproducibility of our findings, each experiment was repeated five times with different random seeds, and the average accuracy across three classes are reported.

\subsection{Algorithms}

Our proposed LSS-CA-SNN was compared with several state-of-the-art approaches, with consistent training procedures and data augmentation:
\begin{enumerate}
\item {\it DeepConvNet and ShallowConvNet}~\cite{Schirrmeister2017}, CNN architectures designed for generic BCI decoding tasks.
\item {\it EEGNet}~\cite{Lawhern2018}, a lightweight CNN for generic BCI decoding tasks. It features a sequence of temporal convolutions followed by spatial depthwise convolutions.
\item {\it EEGConformer}~\cite{Song2022}, which integrates attention mechanisms to enhance the decoding capability of CNN.
\end{enumerate}

Our proposed SpikeDrop was compared with several commonly used data augmentation approaches:
\begin{enumerate}
\item {\it No augmentation} (Baseline), which does not perform any data augmentation at all.
\item {\it Cutmix}~\cite{Yun2019}, which merges segments from two different trials to create a new synthetic trial, enhancing the model's ability to generalize across varying data representations.
\item {\it Mixup}~\cite{Zhang2018b}, which blends two trials in a predefined proportion, generating new data points that help the model learn from interpolated examples.
\item {\it Temporal shift} (TShift)~\cite{Li2022}, which adds a minor offset in the time dimension, simulating variations in the temporal sequence of the data.
\item {\it Temporal reversal} (TReversal)~\cite{Li2022}, which reverses the temporal order of the data, challenging the model to recognize patterns regardless of the sequence direction.
\item {\it Frequency noise} (FNoise)~\cite{Wang2023}, which adds random perturbations in the frequency domain through Fourier transformation, enhancing the model's robustness to frequency-based distortions.
\item {\it Frequency shift} (FShift)~\cite{Wang2023}, which adjusts the frequency components of the data through Fourier transform, simulating changes in the spectral content.
\item {\it NDA}~\cite{Li2022}, which combines image processing techniques such as scrolling, cropping, and cutting, adapted for event-based data.
\item {\it Eventmix}~\cite{Shen2023}, which utilizes gaussian mixture models to generate masks for data mixing, creating diverse synthetic events that broaden the model's exposure to various data shapes.
\end{enumerate}

For augmentations in the frequency domain, data were normalized back to binary following inverse transformation to maintain consistency with the spiking nature of the neural signals. In unsupervised transfer, we set M half the size of the dataset ($M=N/2$), to increase the data pattern of the source session. In supervised transfer, we performed data augmentation on available data in the target session.

\subsection{Main Results}

Tables~\ref{table:combined_motor_paradigm}  and \ref{table:combined_sensory_paradigm} show the classification accuracies of different networks in different scenarios:
\begin{enumerate}
\item EEGNet, ShallowConvNet and DeepConvNet performed relatively well across the motor and sensory paradigms, with EEGNet frequently achieving the best performance among the three in various scenarios.

\item Our proposed LSS-CA-SNN consistently achieved the highest classification accuracies across both paradigms and both scenarios. More specifically, LSS-CA-SNN outperformed EEGNet by 3.87\% and 2.96\% in the motor paradigm for unsupervised transfer and supervised transfer, respectively, and 0.84\% and 0.80\% in the sensory paradigm.
\end{enumerate}

\begin{table*}[htbp]
\caption{Classification accuracies (\%) in the motor paradigm. The best average performance in each panel is marked in bold, and the second best by an underline.} \label{table:combined_motor_paradigm}
\setlength{\tabcolsep}{4pt}  \centering
\begin{tabular}{c|c|cccccccccccccc|c}
\toprule[1pt]
\text{Scenarios} & \text{Approachs} & S0 & S1 & S2 & S3 & S4 & S5 & S6 & S7 & S8 & S9 & S10 & S11 & S12 & S13 & \text{Acc.(\%)} \\ \midrule
\multirow{5}{*}{\text{\shortstack{Unsupervised \\transfer}}} & EEGNet & 82.30 & 86.82 & 87.19 & 72.73 & 93.12 & 90.85 & 95.15 & 89.74 & 77.97 & 78.36 & 87.60 & 81.76 & 67.79 & 63.31 & \underline{82.48}* \\
& ShallowConvNet & 76.90 & 84.66 & 83.60 & 71.09 & 90.15 & 88.21 & 92.61 & 90.36 & 69.52 & 83.03 & 87.28 & 80.13 & 81.90 & 61.63 & 81.51* \\
& DeepConvNet & 83.48 & 83.96 & 81.77 & 64.32 & 92.27 & 88.83 & 94.24 & 88.09 & 78.22 & 73.05 & 84.37 & 82.33 & 70.68 & 60.87 & 80.46* \\
& EEGConformer & 76.80 & 86.23 & 73.02 & 59.18 & 84.81 & 82.30 & 90.92 & 82.53 & 77.03 & 66.23 & 82.89 & 75.99 & 59.51 & 50.39 & 74.84* \\
& LSS-CA-SNN & 86.04 & 91.81 & 91.89 & 80.20 & 94.22 & 93.82 & 96.18 & 93.70 & 75.30 & 86.12 & 90.35 & 86.47 & 73.84 & 68.90 & \textbf{86.35}\hspace{0.5em} \\ \midrule
\multirow{5}{*}{\text{\shortstack{Supervised \\transfer}}} & EEGNet & 85.55 & 90.56 & 89.46 & 81.86 & 94.18 & 93.91 & 94.77 & 95.13 & 88.32 & 90.78 & 90.11 & 84.83 & 87.70 & 81.23 & \underline{89.17}* \\
& ShallowConvNet & 79.42 & 87.61 & 86.78 & 76.97 & 90.37 & 90.91 & 92.95 & 92.24 & 80.48 & 86.40 & 88.45 & 82.43 & 87.67 & 75.22 & 85.56* \\
& DeepConvNet & 84.49 & 90.07 & 88.09 & 78.83 & 92.10 & 90.57 & 94.86 & 92.12 & 86.00 & 84.52 & 87.57 & 83.97 & 87.08 & 78.10 & 87.03* \\
& EEGConformer & 78.78 & 86.59 & 79.66 & 72.30 & 86.66 & 85.72 & 91.63 & 91.25 & 80.10 & 78.82 & 83.95 & 79.87 & 79.47 & 67.63 & 81.60* \\
& LSS-CA-SNN & 89.10 & 94.18 & 94.30 & 85.66 & 95.36 & 96.38 & 95.93 & 96.18 & 89.74 & 91.72 & 93.39 & 88.19 & 93.60 & 86.01 & \textbf{92.13}\hspace{0.5em} \\ \bottomrule[1pt]
\end{tabular}
\vspace{1.0em}
\\{* represents statistically significant difference between LSS-CA-SNN and the corresponding model ($p<0.01$)}.
\end{table*}

\begin{table*}[htbp]
\caption{Classification accuracies (\%) in the sensory paradigm. The best average performance in each panel is marked in bold, and the second best by an underline.} \label{table:combined_sensory_paradigm}
\setlength{\tabcolsep}{5pt} \centering
\begin{tabular}{c|c|cccccccccccc|c}
\toprule[1pt]
\text{Scenarios} & \text{Approachs} & S0 & S1 & S2 & S3 & S4 & S5 & S6 & S7 & S8 & S9 & S10 & S11 & \text{Acc.(\%)} \\ \midrule
\multirow{5}{*}{\text{\shortstack{Unsupervised \\transfer}}} & EEGNet & 74.57 & 63.27 & 96.31 & 97.41 & 97.00 & 97.16 & 95.61 & 97.44 & 94.75 & 95.39 & 94.45 & 89.39 & \underline{91.06}* \\
 & ShallowConvNet & 76.31 & 61.39 & 92.53 & 95.17 & 95.51 & 95.89 & 94.75 & 95.66 & 93.27 & 93.32 & 92.63 & 84.79 & 89.27* \\
 & DeepConvNet & 72.25 & 55.13 & 96.45 & 96.54 & 97.00 & 96.68 & 95.23 & 97.35 & 94.50 & 94.94 & 94.26 & 88.68 & 89.92* \\
 & EEGConformer & 70.30 & 60.17 & 94.53 & 96.11 & 95.60 & 96.37 & 94.61 & 96.86 & 94.51 & 94.44 & 93.46 & 86.60 & 89.46* \\
 & LSS-CA-SNN & 81.83 & 61.39 & 94.86 & 97.91 & 97.83 & 97.24 & 96.75 & 97.13 & 95.03 & 96.43 & 94.90 & 91.48 & \textbf{91.90}\hspace{0.5em} \\ \midrule
\multirow{5}{*}{\text{\shortstack{Supervised \\transfer}}} & EEGNet & 89.50 & 87.85 & 96.62 & 97.43 & 97.22 & 97.15 & 96.05 & 97.03 & 94.87 & 95.69 & 95.15 & 92.65 & \underline{94.77}* \\
 & ShallowConvNet & 83.33 & 83.38 & 93.44 & 95.90 & 95.71 & 96.05 & 94.78 & 95.72 & 93.13 & 94.65 & 92.83 & 87.00 & 92.16* \\
 & DeepConvNet & 89.84 & 86.03 & 96.45 & 97.21 & 97.34 & 96.77 & 95.80 & 97.15 & 94.71 & 95.27 & 95.35 & 91.91 & 94.48* \\
 & EEGConformer & 82.67 & 78.21 & 94.76 & 96.04 & 95.47 & 97.31 & 94.91 & 97.34 & 95.17 & 95.06 & 93.59 & 89.14 & 92.47* \\
 & LSS-CA-SNN & 90.75 & 89.88 & 96.28 & 97.93 & 97.59 & 97.92 & 97.10 & 97.34 & 95.10 & 96.67 & 95.91 & 94.32 & \textbf{95.57}\hspace{0.5em} \\ \bottomrule[1pt]
\end{tabular}
\vspace{1.0em}
\\{* represents statistically significant difference between LSS-CA-SNN and the corresponding model ($p<0.01$)}.
\end{table*}

\begin{figure}     \centering
    \includegraphics[width=0.98\linewidth]{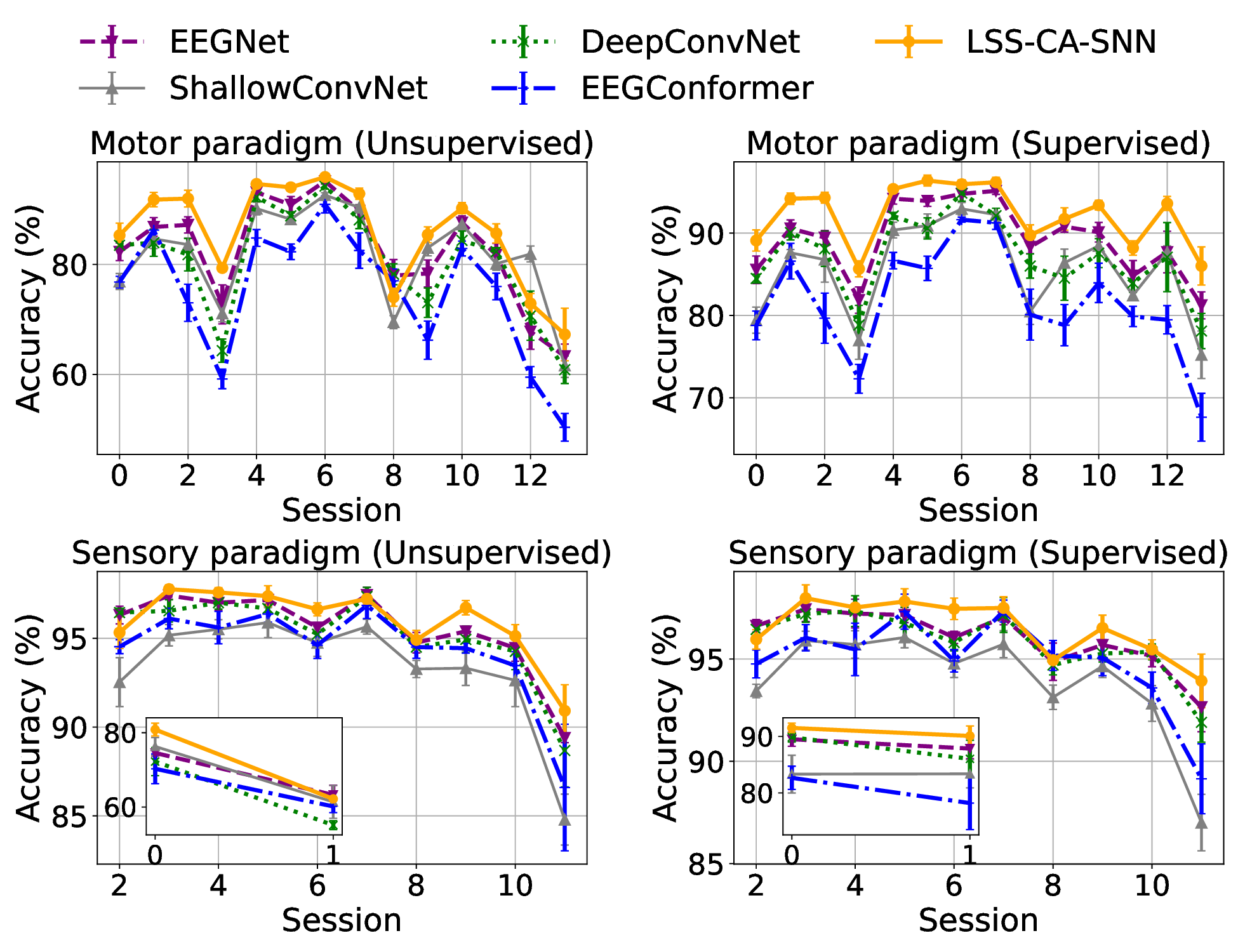}
    \caption{Performance of five neural networks in motor and sensory paradigms (different rows) in two different transfer scenarios (different columns). The error band represents the standard deviation of five repeated experiments.}     \label{fig6result}
\end{figure}

Fig.~\ref{fig6result} shows the results of each network in different scenarios and the standard deviation of repeated experiments. LSS-CA-SNN not only outperformed ANN models in terms of accuracy, but also demonstrated better consistency across various scenarios.

\begin{figure}[htpb]     \centering
    \includegraphics[width=0.98\linewidth,clip]{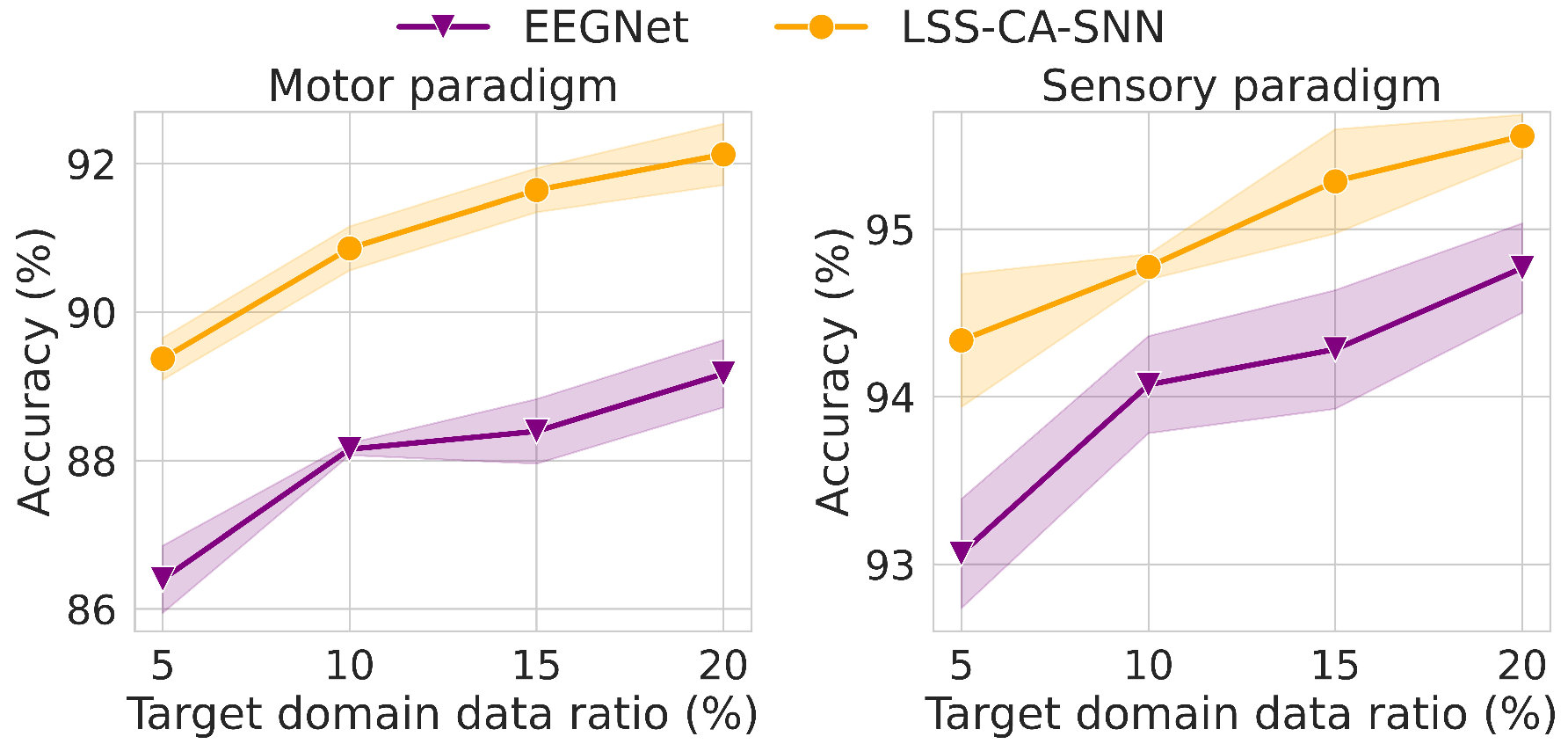}
    \caption{Performance of our proposed LSS-CA-SNN and EEGNet using different target domain data ratios.}     \label{fig7}
\end{figure}

Fig.~\ref{fig7} compares the performance of our proposed LSS-CA-SNN with EEGNet using different target domain data ratios ($5\%$ to $20\%$). LSS-CA-SNN always outperformed EEGNet, i.e., LSS-CA-SNN generalized better and adapted more efficiently to new tasks.

Tables~\ref{table6_B_aug} and \ref{table7_H_aug} show the classification accuracies of various data augmentation approaches in different scenarios:
\begin{enumerate}
\item Image-based augmentations did not perform well on spiking neural data. For instance, in unsupervised transfer of the motor paradigm, both Cutmix and Mixup consistently underperformed all others, with ShallowConvNet's accuracy dropping from 81.51\% (baseline) to 79.47\% (Cutmix) and 80.96\% (Mixup).

\item Time-frequency-based augmentations showed only marginal improvements, and their effectiveness was inconsistent across different models and paradigms. For example, TShift and FShift improved EEGNet's accuracy slightly but failed to boost LSS-CA-SNN's performance.

\item Event-based augmentations showed some promise, particularly in supervised transfer, but their benefits were inconsistent.
	
\item SpikeDrop significantly enhanced the performance of LSS-CA-SNN in both motor and sensory paradigms. For the motor paradigm, LSS-CA-SNN achieved an accuracy of 87.35\% with SpikeDrop, outperforming other augmentations such as NDA (85.97\%). Similarly, in the sensory paradigm, SpikeDrop boosted the accuracy of LSS-CA-SNN to 93.76\%, outperforming NDA's 92.64\%.
	
\item SpikeDrop demonstrated consistent improvements across diverse architectures and paradigms, confirming its broad applicability. For example, SpikeDrop improved the accuracy of EEGNet from 82.48\% to 83.98\% in the motor paradigm in unsupervised transfer, and improved the accuracy of EEGConformer from 92.47\% to 93.34\% in the sensory paradigm in supervised transfer.
\end{enumerate}

\begin{table*}[htbp]
\caption{Classification accuracies (\%) in the motor paradigm using different data augmentations. The best average performance in each approach is marked in bold, and the second best by an underline.} \label{table6_B_aug} \setlength{\tabcolsep}{5pt} \centering
\begin{tabular}{c|c|cccccccccc}
\toprule[1pt]
Scenario & Approach & {Baseline} & {Cutmix} & {Mixup} & {TShift} & {TReversal} & {FNoise} & {FShift} & {Eventmix} & {NDA} & {SpikeDrop (ours)} \\ \midrule
\multicolumn{1}{c|}{\multirow{5}{*}{\shortstack{Unsupervised \\ transfer}}} & EEGNet & 82.48 & 80.89 & 82.26 & \underline{82.85} & 80.99 & 82.11 & { 82.83} & 82.26 & 82.09  & {\bf 83.98} \\
 & ShallowConvNet & {\bf 81.51} & 79.47 & 80.96 & 81.12 & 80.42 & 81.15 & 81.13 & 80.96 & 81.11  & \underline{81.17} \\
 & DeepConvNet & 80.46 & 79.27 & 80.78 & 81.18 & 79.05 & 80.24 & \underline{81.87} & 80.78 & 80.92  & {\bf 83.04} \\
 & EEGConformer & 74.84 & 75.13 & 75.27 & \underline{75.91} & 73.55 & 75.30 & {75.84} & 75.27 & 75.29  & {\bf 77.04} \\
 & LSS-CA-SNN (ours) & \underline{86.35} & 83.20 & {86.32} & 85.88 & 83.85 & 86.10 & 85.42 & {86.32} & 85.97  & {\bf 87.35} \\ \midrule
\multicolumn{1}{c|}{\multirow{5}{*}{\shortstack{Supervised \\ transfer}}} & EEGNet & 89.17 & 86.85 & 89.27 & \underline{89.76} & 88.63 & 89.45 & 89.52 & 89.27 & 89.64 & {\bf 89.85} \\
 & ShallowConvNet & 85.56 & 82.76 & 85.85 & 85.80 & 84.86 & 85.50 & 85.30 & 85.85 & \underline{85.89} & {\bf 85.90} \\
 & DeepConvNet & 87.03 & 84.31 & 87.03 & 87.52 & \underline{87.67} & { 87.59} & {\bf 88.31} & 87.03 & 87.38 & 87.41 \\
 & EEGConformer & 81.60 & 80.34 & \underline{82.19} & 82.11 & 81.09 & 81.63 & 81.51 & \underline{82.19} & 82.03 & {\bf 82.43} \\
 & LSS-CA-SNN (ours) & 92.13 & 89.16 & 92.18 & 92.26 & 91.31 & \underline{92.45} & 91.97 & 92.18 & 92.44 & {\bf 92.50} \\ \bottomrule[1pt]
\end{tabular}
\end{table*}

\begin{table*}[htbp]
\caption{Classification accuracies (\%) in the sensory paradigm using different data augmentations. The best average performance in each approach is marked in bold, and the second best by an underline.} \label{table7_H_aug} \setlength{\tabcolsep}{5pt} \centering
\begin{tabular}{c|c|cccccccccc}
\toprule[1pt]
Scenario & Approach & Baseline & Cutmix & Mixup & TShift & TReversal & FNoise & FShift & Eventmix & NDA & SpikeDrop (ours) \\ \midrule
\multicolumn{1}{c|}{\multirow{5}{*}{\shortstack{Unsupervised \\ transfer}}} & EEGNet & 91.06 & \underline{91.68} & 90.76 & 90.85 & 89.42 & 90.90 & 91.36 & 90.76 & 90.84 & \textbf{92.37} \\
& ShallowConvNet & 89.27 & 88.87 & 88.96 & 89.37 & 88.09 & \underline{89.47} & 89.18 & 88.96 & 88.92 & \textbf{90.19} \\
 & DeepConvNet & 89.92 & {90.57} & 90.17 & 90.23 & 89.17 & \underline{90.58} & 90.33 & 90.17 & 90.34 & \textbf{92.19} \\
 & EEGConformer & 89.46 & \underline{90.49} & 89.05 & 89.24 & 87.04 & 89.12 & 89.77 & 89.05 & 89.05 & \textbf{90.77} \\
 & LSS-CA-SNN (ours) & {91.90} & 91.28 & 91.84 & 91.90 & 91.11 & 92.25 & 91.68 & 91.42 & \underline{92.64} & \textbf{93.76} \\ \midrule
\multicolumn{1}{c|}{\multirow{5}{*}{\shortstack{Supervised \\ transfer}}} & EEGNet & 94.77 & 94.98 & \underline{95.00} & \textbf{95.01} & 94.70 & 94.98 & 94.96 & {94.99} & 94.89 & {94.99} \\
 & ShallowConvNet & 92.16 &  \underline{92.37} & 92.30 & 92.09 & 92.22 & 91.94 & 92.19 & 92.30 & \textbf{92.54} & {92.32} \\
 & DeepConvNet & 94.48 & 94.58 & \underline{94.75} & {94.62} & 94.38 & 94.51 & 94.40 & \textbf{94.76} & 94.42 & {94.37} \\
 & EEGConformer & 92.47 & 92.82 & \underline{92.93} & 92.46 & 92.33 & 92.61 & 92.64 & {92.91} & 92.78 & \textbf{93.34} \\
 & LSS-CA-SNN (ours) & {95.57} & 95.55 & 95.52 & 95.48 & \underline{95.60} & 95.46 & 95.47 & 95.34 & 95.30 & \textbf{95.64} \\ \bottomrule[1pt]
\end{tabular}
\end{table*}

In summary, our proposed LSS-CA-SNN achieved state-of-the-art classification performance, with good generalization and consistency across different transfer scenarios and paradigms. SpikeDrop, specifically designed to preserve the discrete and temporal characteristics of spiking data, can be used to further improve the performance of LSS-CA-SNN and traditional ANNs.

\subsection{Energy Consumption Comparison}

ANNs primarily rely on continuous signal propagation and floating point operations (FLOPs) measured in multiply-accumulate (MAC) operations, whereas SNNs utilize event driven discrete accumulate (AC) operations to calculate the FLOPs. In a generic task for SNNs, the initial encoding layer performs MAC operations for real-valued inputs, and subsequent layers execute AC operations. In invasive BCIs, the first convolutional layer is transformed into an AC operation due to the discrete nature of spiking inputs.

Let $E {vanilla}$ be the energy consumption of a standard SNN model. For spiking data in binary form, the energy consumption can be quantified as:
\begin{align}
E_{vanilla}=E_{AC}\cdot\left(\sum_{n=1}^{N}FLOPs_{Conv_n}+\sum_{m=1}^{M}FLOPs_{FC_m}\right),
\label{energy consumption2}
\end{align}
where $N$ and $M$ represent the total number of convolutional and fully connected layers, respectively, and $E_{MAC}$ and $E_{AC}$ respectively the energy consumption of MAC and AC operations.

As in prior research~\cite{Yao2023}, we assume 32-bit floating-point arithmetic implemented using 45 nm technology, where the energy consumption for an MAC operation ($E_{MAC}$) is 4.6 pJ, and for an AC operation ($E_{AC}$) is 0.9 pJ.

The channel-wise attention module reduces energy consumption. The absolute energy shift, $\Delta E$, is:
\begin{align}
    \Delta_{E}=E_{MAC} \cdot \Delta_{MAC}-E_{AC} \cdot \Delta_{AC}, \label{absolate energy shift}
\end{align}
where $\Delta_{MAC}$ is the additional MAC operations, and $\Delta_{AC}$ the reduced AC operations. The energy consumption of attention-based SNN can be expressed as $E_{Att}$, and the energy efficiency ratio $r_{EE}$ is defined by comparing it with $E_{vanilla}$:
\begin{align}
    r_{EE}=\frac{E_{vanilla}}{E_{Att}}= \frac{E_{vanilla}}{E_{vanilla}+\Delta _E}.
\label{energy consumtion ratio}
\end{align}

We also introduce spiking counts as a direct measure of neuronal activities:
\begin{align}
    SC = N_l\cdot T_l \cdot fr,
\end{align}
where $N_l$ denotes the number of neurons in layer $l$, $T_l$ the time steps of layer $l$, and $fr$ the spiking activity rate. Note that the source of AC operations in SNN is the activation of neurons, and fewer spiking counts mean fewer AC operations and less energy consumption.

Additionally, we introduce the network average spiking activity rate (NASAR)~\cite{Yao2023} as a measure of energy consumption variation within the network. NASAR provides a straightforward metric for tracking changes in energy consumption. While $r_{EE}$ offers a detailed evaluation of energy shifts, NASAR serves as a simpler method for monitoring these variations, with spiking counts providing a direct observation of neuronal activities.

Table~\ref{table9allresult} shows energy consumption of different approaches.

\begin{table*}[hbtp]
\caption{Average accuracies (\%) and energy consumptions (uJ) of different approaches. The number in parentheses shows the efficiency ratio relative to LSS-CA-SNN in the same scenario.} \label{table9allresult}
\setlength{\tabcolsep}{7pt} \centering
\begin{tabular}{@{}cccccccccc@{}}
\toprule[1pt]
\multirow{4}{*}{Scenario} & \multirow{4}{*}{Approach} & \multicolumn{4}{c}{Motor paradigm} & \multicolumn{4}{c}{Sensory paradigm} \\ \cmidrule(l){3-6} \cmidrule(l){7-10}
 &  & \multirow{2}{*}{\begin{tabular}[c]{@{}c@{}}Average\\ accuracy (\%)\end{tabular}} & \multicolumn{2}{c}{FLOPs (M)} & \multirow{2}{*}{\begin{tabular}[c]{@{}c@{}}Energy \\ consumption (uJ)\end{tabular}} & \multirow{2}{*}{\begin{tabular}[c]{@{}c@{}}Average\\ accuracy (\%)\end{tabular}} & \multicolumn{2}{c}{FLOPs (M)} & \multirow{2}{*}{\begin{tabular}[c]{@{}c@{}}Energy \\ consumption (uJ)\end{tabular}} \\ \cmidrule(lr){4-5} \cmidrule(lr){8-9}
 &  &  & MAC & AC &  &  & MAC & AC &  \\ \midrule
\multirow{5}{*}{\begin{tabular}[c]{@{}c@{}}Unsupervised\\ transfer\end{tabular}} & EEGNet & \underline{82.48} & 4.37 & 0 & \underline{20.10 (14.89$\times$)} & \underline{91.06} & 3.61 & 0 & \underline{16.61 (31.94$\times$)} \\
 & ShallowConvNet & 81.51 & 16.06 & 0 & 73.88 (54.73$\times$) & 89.27 & 13.25 & 0 & 60.95  (117.21$\times$) \\
 & DeepConvNet & 80.46 & 6.80 & 0 & 31.28 (23.17$\times$) & 89.92 & 5.76 & 0 & 26.50 (50.96$\times$) \\
 & EEGConformer & 74.84 & 16.21 & 0 & 74.57 (55.24$\times$) & 89.46 & 13.40 & 0 & 61.64 (118.54$\times$) \\
 & LSS-CA-SNN (ours) & \textbf{86.35} & 0.03 & 1.33 & \textbf{1.35 (1$\times$)} & \textbf{91.90} & 0.03 & 0.44 & \textbf{0.52 (1$\times$)} \\ \midrule
\multirow{5}{*}{\begin{tabular}[c]{@{}c@{}}Supervised\\ transfer\end{tabular}} & EEGNet & \underline{89.17} & 4.37 & 0 & \underline{20.10 (14.78$\times$)} & \underline{94.77} & 3.61 & 0 & \underline{16.61 (31.34$\times$)} \\
 & ShallowConvNet & 85.56 & 16.06 & 0 & 73.88 (54.32$\times$) & 92.16 & 13.25 & 0 & 60.95  (115.01$\times$) \\
 & DeepConvNet & 87.03 & 6.80 & 0 & 31.28 (23.17$\times$) & 94.48 & 5.76 & 0 & 26.50 (50.01$\times$) \\
 & EEGConformer & 81.60 & 16.21 & 0 & 74.57 (54.83$\times$) & 92.47 & 13.40 & 0 & 61.64 (116.30$\times$) \\
 & LSS-CA-SNN (ours) & \textbf{92.13} & 0.03 & 1.34 & \textbf{1.36 (1$\times$)} & \textbf{95.57} & 0.03 & 0.45 & \textbf{0.53 (1$\times$)} \\ \bottomrule[1pt]
\end{tabular}
\end{table*}

In the motor paradigm, LSS-CA-SNN was the best performer, achieving an average accuracy of 86.35\% with an energy consumption of only 1.35 uJ. Compared to EEGNet with an accuracy of 82.48\% and energy consumption of 20.10 uJ, LSS-CA-SNN improved the accuracy by 3.87\% and energy efficiency 14.89 times. The performance improvements over ShallowConvNet and DeepConvNet were more significant.

In the sensory paradigm, although the improvements in classification accuracies were not large, there were still significant energy efficiency gains. LSS-CA-SNN achieved an average accuracy of 91.90\%, with an energy consumption of 0.52 uJ, 31.94-118.54 times more efficient than other networks.

In summary, LSS-CA-SNN excelled in both classification accuracy and energy efficiency in the motor paradigm, and also achieved substantial energy savings in the sensory paradigm, indicating that LSS-CA-SNN is promising for energy-sensitive applications, e.g., portable devices and real-time edge computing.

\subsection{Ablation Studies}

Table~\ref{table10ablation} shows ablation study results in unsupervised transfer of the motor paradigm. While LSS or CA alone was beneficial, combining them achieved the best performance in both classification accuracy and energy efficiency.

\begin{table}[hbtp]
	\caption{Effects of LSS and CA on the classification performance and energy consumption.} 	\label{table10ablation}
	\setlength{\tabcolsep}{3pt}	\centering
	\begin{tabular}{@{}cc|ccccccc@{}}
		\toprule[1pt]
		\multicolumn{2}{c|}{Module} & \multirow{2}{*}{Acc. (\%)} & \multirow{2}{*}{$r_{EE}$} & \multirow{2}{*}{\shortstack{Spiking \\ counts}} & \multicolumn{4}{c}{NASAR}  \\ \cmidrule(r){1-2} \cmidrule(lr){6-9}
		LSS & CA &  &  &  & {\footnotesize layer 1} & {\footnotesize layer 2} & {\footnotesize layer 3} & {\footnotesize layer 4}   \\ \midrule
		--&--  & 85.06 & 1$\times$ & 136 & 0.107 & 0.024 & 0.209 & 0.286  \\
		$\surd$  & -- & 85.64 & 1.09$\times$ & 131 (-3.7\%) & 0.104 & 0.022 & 0.204 & 0.231 \\
		--& $\surd$  & 85.92 & 1.10$\times$ & 132 (-2.9\%) & 0.117 & 0.022 & 0.188 & 0.165 \\
		$\surd$  & $\surd$  & {\bf 86.35} & {\bf 1.12$\times$} & {\bf 109 (-19.9\%)} & 0.099 & 0.019 & 0.115 & 0.048 \\ \bottomrule[1pt]
	\end{tabular}
\end{table}

\section{Conclusions} \label{sec:conclusion}

This study has introduced LSS-CA-SNN, a novel SNN architecture that integrates LSS and CA for intracortical spiking signal decoding. LSS optimizes the neuronal membrane potential and improves the classification performance, and CA optimizes neuronal activation and reduces energy consumption. LSS-CA-SNN outperformed traditional ANNs in terms of both classification accuracy and energy consumption in two different paradigms. We also proposed SpikeDrop, a spiking data augmentation approach to further enhance model generalization. Our findings highlight the potential of SNNs in decoding cortical signals with high accuracy and efficiency.


\end{document}